\documentclass[twocolumn,showpacs,preprintnumbers,amssymb]{revtex4}
\usepackage{graphicx}
\usepackage{dcolumn}
\usepackage{amsmath}
\usepackage{amssymb}
\begin{document}

\title{The scale-free texture of the fast solar wind.}
\author{B. Hnat$^{1}$, S. C. Chapman$^{1}$, G. Gogoberidze$^{1,2}$, R. T. Wicks$^{3}$}
\affiliation{$^{1}$Centre for Fusion, Space and Astrophysics, Physics Department, University of Warwick, Coventry, UK\\
$^{2}$Institute of Theoretical Physics, Ilia State University, 3/5 Cholokashvili Ave., 0162 Tbilisi, Georgia\\
$^{3}$Space and Atmospheric Group, Physics Department, Imperial College, UK}
\date{\today} % It is always today, date may be explicitly specified

\begin{abstract}
The higher order statistics of magnetic field magnitude fluctuations in the fast quiet solar wind are quantified
systematically scale-by-scale for the first time. We find a single global non-Gaussian scale free
behaviour from minutes to over $5$ hours. This spans the signature of an inertial range of
magnetohydrodynamic (MHD) turbulence and a $\sim \!\! 1/f$ range in magnetic field components. This global scaling
in field magnitude fluctuations is an intrinsic component of the underlying texture of the solar wind and it
suggests a single stochastic  process for magnetic field magnitude fluctuations operating across the full range
of MHD time scales supported by the solar wind. Intriguingly, the magnetic field and velocity components show scale
dependent dynamic alignment outside of the inertial range.
\end{abstract}

\maketitle
In-situ spacecraft observations of plasma parameters in the interplanetary high Reynolds number solar wind flow \cite{matt07}
provide time series over several decades that are ideally suited to studies of turbulence and of other phenomena that generate
statistical scaling. These observations show a power law range of power spectral density (PSD) in magnetic field and velocity
components which is associated with an inertial range of magnetohydrodynamic (MHD) turbulence (see e.g.\cite{GRM}).
This inertial range phenomenology is seen from ion kinetic scales up to an outer scale at which there is a cross-over to
a $\sim \!\! 1/f$  power law range of the PSD in the field components of coronal origin \cite{matthaeus_86,matt07}.

There is debate as to which aspects of these observations should be attributed to MHD turbulence. The dominant velocity
and magnetic field components fluctuations are found in the inertial range of scales and these {\em do} exhibit
statistical signatures consistent with evolving MHD turbulence: intermittency \cite{horbury}, non-Gaussian statistics
\cite{marsch_97} and anisotropy \cite{milanoprl,chapmangrl}. This inertial range extends to lower frequencies both with
increasing solar distance and in slow as compared to fast solar wind suggesting an actively evolving turbulent cascade
\cite{horbury_96b,marschtuspecevolv}. However, such a cascade evolves in the presence of fluctuations of coronal origin,
the `texture' of the solar wind \cite{brunospagettons,borovsky}. This non-trivial texture, in the
absence of turbulence, can be inferred, at least on longer scales, from the observed $\sim \!\! 1/f$ fluctuations in the
magnetic field components. These show latitudinal variation and statistical scaling distinct from that of the inertial range
\cite{matt07,nicolapjlong,ESSprl}.

Since the earliest in-situ observations \cite{coleman} the relative importance of turbulence in these signatures has been
debated. Recently, it has been argued that the range of PSD scaling exponents routinely measured in the inertial range for
magnetic field components can arise from a non-evolving set of discontinuities \cite{borovskyPRL}. However, distinct physical
mechanisms can share the same power law exponent in the PSD \cite{npg}. In the context of Alfv\'{e}nic MHD turbulence in
the solar wind, where discussion naturally centres on the components of velocity and magnetic field, such discontinuities
can arise from turbulence or be part of the solar wind texture. There is no unique procedure to distinguish these two physical
signatures. To understand the texture of the solar wind it is thus informative to consider scalar parameters alongside vector
components.

Intriguingly, in the solar wind, scalar plasma parameters such as magnetic field magnitude $|{\bf B}|$, energy density $B^2$
and density $n$ show  scaling and this is also a feature of the texture of the solar wind. A power law range in the  PSD of 
$|{\bf B}|$ is seen to extend through  the inertial range into the $\sim \!\! 1/f$ range of frequencies but at $1$ AU is typically
a decade lower in power compared to that of the magnetic field components $B_k$ \cite{goldrobR}. Scalar magnetic field and plasma
parameters can also show scaling in long time bulk hourly averages \cite{burlagamag} and aggregate statistics up to tens of hours
\cite{hnatgrl02,kiyani07}. Pioneering work with HELIOS \cite{denk1982,bav,marschtu1990} showed that $|{\bf B}|$ power
spectral exponent is roughly invariant with radial distance at frequencies below $10^{-2} $Hz; a `flattening' at higher
frequencies seen in the inner heliosphere is not seen at 1AU (however our analysis will not approach these high frequencies).
There is an admixture of compressive fluctuations and pressure balanced structures \cite{tumarsch1994}; $|{\bf B}|$ and $n$ 
do not simply advect together as passive scalars \cite{hnatprl}.

Distinct physical mechanisms that generate scaling can share the same power law exponent in the PSD \cite{npg} and thus statistical
scaling of the higher order moments is an essential tool needed to distinguish them. 
In this Letter we perform the first systematic scale-by-scale study of the statistical scaling properties of $|{\bf B}|$
fluctuations in extended intervals of fast, quiet solar wind. We find global scale-free behaviour in $|{\bf B}|$ fluctuations
which occurs alongside, but is quite distinct from, that of the components of magnetic field, and extends through both the
inertial and $\sim \!\! 1/f$ ranges of temporal scales. Thus we establish scaling is an intrinsic feature of the texture of the solar
wind operating through the inertial and $\sim \!\! 1/f$ ranges. This suggests that a single stochastic  process for magnetic field
magnitude fluctuations is operating or has operated across this full range of MHD time scales supported by the solar wind.

A corollary of this result is that it provides a natural laboratory to test proposed measures of in-situ MHD turbulence, since
within the same dataset fluctuations may be generated by turbulence (the inertial range) and on longer scales may be of solar
origin. Scaling exponents predicted by theories of turbulence are difficult to determine accurately in data \cite{deWit04,Kiyani09},
hence the attraction to test signatures such as scale dependent dynamic alignment \cite{podesta}. Direct numerical simulations
(DNS) \cite{Muller,Beresnyak} have suggested such alignment \cite{Boldyrev95}, in which turbulent fluctuations of velocity and
magnetic field progressively align in the cascade, to be a signature of anisotropic MHD turbulence. This has recently attracted 
controversy \cite{lazarian,Beres11}, which we will address directly from the observations.

We use in-situ observations of fast solar wind from the ACE and Ulysses spacecraft (data obtained from the CDAWeb site
http://cdaweb.gsfc.nasa.gov/). ACE is at $\sim \!\! 1$AU in the ecliptic, whereas for the intervals under study, Ulysses was at a
heliospheric latitude above $70\,^{\circ}$ at $\sim2.2~{\rm AU}$; thus we can compare solar wind of distinct coronal origin.
We selected continuous intervals of quiet fast solar wind flow at solar minimum which did not contain large coherent structures or
large scale, long time secular change in the ion plasma pressure (estimated as $nT$). Fast wind intervals have been identified using
the following criteria: $\left< v_{sw} \right> \geq 550~{\rm km/s}$ and its standard deviation $\sigma (v_{sw}) \leq 50~{\rm km/s}$,
where $v_{sw}$ is the speed of the solar wind. Five stationary fast solar wind intervals were identified in ACE $2007$ and $2008$
datasets of $64$ second average calibrated magnetic field magnitude and plasma observations. The Ulysses observations correspond to
$5$ intervals of $3$ days duration (July $1$-$17$) of the $1995$ (solar minimum) north polar pass. We used $1$ minute average calibrated
magnetic field data for the magnetic field analysis only. Ulysses plasma $4$ minutes averaged observations were combined with magnetic
field data by selecting the $1$ second cadence magnetic field records that match plasma observations times to the nearest second.

All statistical measures are first computed individually for each interval, these are cubic spline interpolated onto a common temporal
or frequency grid and then an average over all $5$ intervals for each of ACE and Ulysses is obtained. The longest time scale is fixed
by that of the shortest data interval so that we can access time scales over the range $\sim \!\! 2$ minutes to $\sim \!\! 20$ hours.
Practically, given the effect of averaging onto $64$ seconds and non-uniformities in the time base, we do not draw strong conclusions
from analysis on time scales shorter than $\sim \!\! 3$ minutes. We will indicate  with error bars on our plots the r.m.s. variation
about the sample average across the intervals.
\begin{figure}[]
\begin{center}
\includegraphics[width=1\columnwidth]{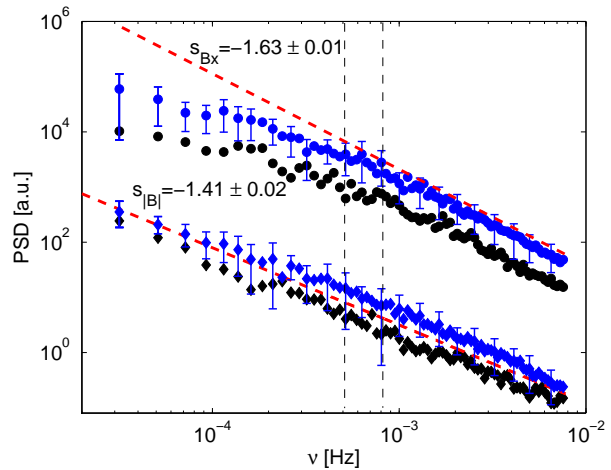}
\caption{Averaged PSD of magnetic field GSE component $B_x$ from ACE (black circle), RTN component $B_R$ from Ulysses (blue circle)
and magnetic field magnitude: ACE-black diamond, Ulysses-blue diamond. PSDs were displaced vertically for clarity; dashed
lines indicate time scales of $20$ and $30$ minutes.}
\label{fig1}
\end{center}
\end{figure}
The averaged, smoothed PSD (Welch modified periodogram) of the magnetic field magnitude for both ACE and Ulysses intervals
is shown in Figure \ref{fig1} alongside that of one of the components: $B_x$ (ACE) and $B_R$ (Ulysses). These curves have been
displaced vertically in the figure for clarity. Typically, the magnetic field components are a factor of $5$ higher in
power compared to $|{\bf B}|$ for these intervals and on this plot show a steeper power-law trend at higher
frequencies with exponent, $s$, close to the often observed $\approx \frac{5}{3}$ `inertial range' value. Figure \ref{fig1} shows
that at a time scale $20$-$30$ minutes the components depart from $\approx \!\! \frac{5}{3}$ inertial range
scaling and cross-over to a $\sim \!\! 1/f$ spectrum. This time scale is indicated on all subsequent plots. The PSD of $|{\bf B}|$
follows a single power-law over the entire range with exponent $\approx \!\! -1.4$, a behaviour that is distinct from that of
the components. The Ulysses and ACE observations show essentially the same behaviour in the PSD: a transition from inertial
range to $\sim \!\! 1/f$ scaling in the components (also seen in the other components, not shown) and alongside this,
a single range of scaling in the magnitude with exponent distinct from that of the inertial range.
\begin{figure}[]
\begin{center}
\includegraphics[width=1\columnwidth]{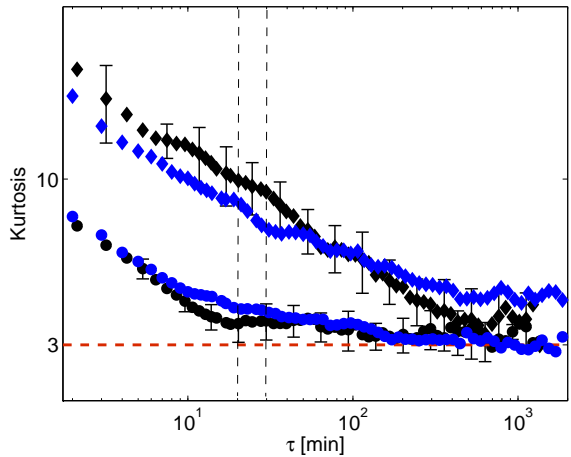}
\caption{Kurtosis of magnetic field component $B_x$ from ACE (black circle), component $B_R$ from Ulysses (blue circle) and magnetic field magnitude: ACE-black diamond, Ulysses-blue diamond.}
\label{fig2}
\end{center}
\end{figure}
We now consider the statistical scaling in the higher moments of the fluctuations. On a given temporal scale $\tau$, the 
fluctuations are $\delta x(t,\tau)=x(t+\tau)-x(t)$ in the time series of a given (scalar or vector) quantity $x(t)$.
In Figure \ref{fig2} we plot, for the same quantities shown in Figure \ref{fig1}, the Kurtosis of the fluctuations
$K=\mu_4/\mu_2^2$ where $\mu_k=<(\delta x(t,\tau)-< \delta x(t,\tau)>)^k>$ are their central moments. For a Gaussian PDF
$K=3$ and this is indicated with a dashed horizontal line. The components (within variance of the average) show a smooth and
scale free drop in Kurtosis within the inertial range, on scales shorter than $\sim \!\! 20-30$ minutes. At $\sim 20-30$ minutes
there is a 'kink' at the end if the inertial range, beyond which the PDFs are weakly non-Gaussian ($K>3$) and progressively
approach Gaussian at scales $\sim 120$ minutes. The behaviour of $\delta |{\bf B}|$ is clearly distinct from that of the
components, and is far from Gaussian until much longer time scales, beyond $\sim \!\! 5$ hours.
On scales shorter than $\sim \!\! 5$  hours the behaviour in the Kurtosis seen in fast quiet solar wind streams in the ecliptic
(ACE) roughly tracks that seen in polar outflows (Ulysses); fluctuations of $|{\bf B}|$ are not sensitive to the distinct coronal 
origin of these flows until the longest temporal scales are reached. We note that, in principle, more information can be obtained
from signed structure functions of odd moments, however, consistent with \cite{carboneprl09,comment}, we find that these signed
quantities do not produce homogeneous results across all intervals studied here, therefore it is not appropriate to average these.

\begin{figure}[]
\begin{center}
\includegraphics[width=1\columnwidth]{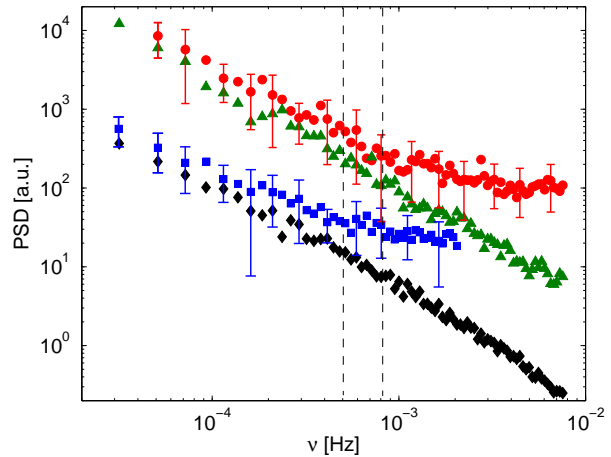}
\caption{Average PSD of magnetic field magnitude (ACE-green triangles, Ulysses-black diamond) and density (ACE -red circle,
Ulysses-blue square).}
\label{fig3}
\end{center}
\end{figure}

Figures \ref{fig1} and \ref{fig2} together establish a quite remarkable result: there is a single scale-free behaviour in 
fluctuations of $|{\bf B}|$ spanning three decades in temporal scales from $\sim \!2$ minutes to over $5$ hours.
The scale-free signatures extend through both the inertial and $\sim \!\! 1/f$ ranges of temporal scales seen in these same intervals
in the field components. This suggests a physical mechanism for $|{\bf B}|$ scaling across all these scales that is distinct from
the in-situ Aflv\'{e}nic MHD turbulence that is driving scaling in the components. Compressive fluctuations and pressure balanced
structures \cite{tumarsch1994} are natural candidates for magnetic field magnitude fluctuations--they can in principle be
dynamically evolving or passively advecting, having been generated in the formation of the solar wind. In Figure \ref{fig3} we
directly compare the PSD of $|{\bf B}|$ from ACE and Ulysses as shown in Figure \ref{fig1} with that of the density.
The scaling of the density closely follows that of $|{\bf B}|$ in the $\sim \!\! 1/f$ range, but departs on time scales shorter
than $20$-$30$ minutes, that is, where the  inertial range is seen in the magnetic field components.
In the inertial range, the power in the density fluctuations is enhanced above that which would arise from the scaling seen on
longer time scales. Essentially the same behaviour is seen in both ACE and Ulysses and thus again is not strongly sensitive to the
region of coronal origin. The single range of scaling in $|{\bf B}|$ fluctuations that extends through both the inertial and
$\sim \!\! 1/f$ ranges of scale thus does not straightforwardly correspond to the behaviour of the density fluctuations. Thus one
cannot invoke a single process to generate the scaling of fluctuations in both $|{\bf B}|$ and density unless some physical
process operating in the inertial range acts to enhance density fluctuations.

\begin{figure}[]
\begin{center}
\includegraphics[width=1\columnwidth]{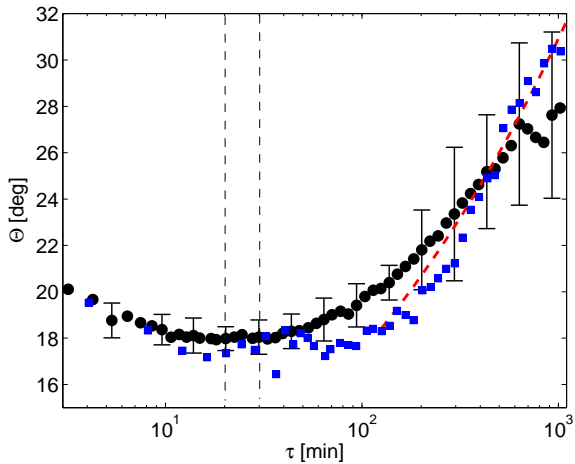}
\caption{Average angle of perpendicular fluctuations in velocity and magnetic field ACE-black diamond, Ulysses-blue square.
An asymptote of $\Theta \sim \tau^{0.25}$ is indicated on the plot.}
\label{fig4}
\end{center}
\end{figure}

Finally, we will look more closely at the behaviour of the components of fluctuations across both the inertial and $\sim \!\! 1/f$
ranges of scale. We explore this idea by performing a dynamic alignment analysis across the full range of scales available in
our chosen intervals. Dynamic alignment, calculated scale-by-scale as the relative angle between the vector fluctuations
${\bf \delta v}_{\perp}$ and ${\bf \delta b}_{\perp}$ \cite{Boldyrev95,podesta}, has been proposed as a signature of the turbulent
cascade \cite{Boldyrev95}.
Perpendicular directions are taken with respect to the mean local magnetic field, which at time $t$ is given by an average
$\left< {\bf B}(t) \right>$ computed over a scale dependent interval $[t-\tau, t+\tau]$. Perpendicular fluctuations are then 
obtained from the following expressions:
${\bf \delta v}_{\perp}={\bf \delta v} - ({\bf \delta v} \cdot \bf{\hat b}) \bf{\hat b}$ and
${\bf \delta b}_{\perp}={\bf \delta b} - ({\bf \delta b} \cdot \bf{\hat b}) \bf{\hat b}$, where the unit vector
$\bf{\hat b}=\bf {\delta B}/B$.
The average angle of alignment is:
\begin{equation}
\label{theta}
\Theta(\tau)=arcsin \left( \frac{\left\langle \mid{\bf \delta v}_{\perp} \wedge {\bf \delta b}_{\perp}\mid \right\rangle} {\left\langle \delta v_{\perp} \right\rangle \left\langle \delta b_{\perp} \right\rangle} \right).
\end{equation}
In Figure \ref{fig4} we plot this average angle $\Theta(\tau)$ as a function of scale $\tau$ for the same intervals shown in
Figures \ref{fig1}-\ref{fig3}. On such a plot, progressively aligning field and flow fluctuations should show a scale free
decrease in $\langle \Theta \rangle$ with decreasing scale $\tau$ within the inertial range; as the cascade progresses from large
to small scales the field and flow progressively align \cite{Boldyrev95}. However, Figure \ref{fig4} show this behaviour most
strongly on temporal scales $\tau>100$ minutes, well into the $\sim \!\! 1/f$ range of scales. The alignment flattens on shorter
scales, reaching its minimum at the upper end of the inertial range at $20$-$30$ minutes. Indeed, the variation of
$\langle \Theta \rangle$ with decreasing scale $\tau$ asymptotes to within errors of that predicted for in-situ turbulence
\cite{Boldyrev95}. Thus ``dynamic alignment", as quantified by (\ref{theta}), is not a clear discriminator of turbulence in the
solar wind, but intriguingly, in the absence of in-situ MHD turbulence, that is, in the $\sim \!\! 1/f$ range it is coincident with
scaling in fluctuations of $|{\bf B}|$.

We have established that there is a scale-free behaviour in $|{\bf B}|$ fluctuations that occurs alongside, but is quite distinct
from, that of the components of magnetic field. It extends over three decades through both the inertial and $\sim \!\! 1/f$
ranges of temporal scales and is seen in fast quiet solar wind of different coronal origin. The most parsimonious description
of how such a single scaling range in $|{\bf B}|$ could arise is that a single process operates, or has operated, over all these
scales. Observational evidence of incompressible MHD turbulence in the solar wind must thus be understood in the context of the
global \textit{scaling} of the `texture' \cite{brunospagettons,borovsky} of the solar wind. 
We have also shown that scale dependent `dynamic alignment' is not a clear discriminator of turbulence in the solar wind.
This `pseudo-dynamic alignment', taken alongside the scaling seen in the magnitude of magnetic field, may however provide an insight
into the generation of the flux tube texture of the solar wind. It may capture some physics of the generation of the solar wind,
reflecting the manner in which magnetic helicity is injected via photospheric fields that are fractal \cite{fsun}.

\acknowledgments
We acknowledge the ACE team for data provision. This work was supported by the UK STFC.

\end{document}